\documentclass[12pt,preprint]{aastex}

\newcommand{\msec}[2]{$#1\mbox{$''\mskip-7.6mu.\,$}#2$}

\newcommand{\sbeamp}[5]{{$#1\mbox{$''\mskip-7.6mu.\,$}#2$} $\times$ {$#3\mbox{$''\mskip-7.6mu.\,$}#4$}; $+#5^\circ$}
\newcommand{\sbeamm}[5]{{$#1\mbox{$''\mskip-7.6mu.\,$}#2$} $\times$ {$#3\mbox{$''\mskip-7.6mu.\,$}#4$}; $-#5^\circ$}

\newcommand{\Msun}{M$_{\odot}$}

\begin{document}

\title{New radio sources and the composite structure of component B\\
      in the very young protostellar system IRAS~16293--2422\\}

\author{Laurent Loinard}
\affil{Centro de Radiostronom\'{\i}a y Astrof\'{\i}sica, 
       Universidad Nacional Aut\'onoma de M\'exico,\\
       Apartado Postal 72--3 (Xangari), 58089 Morelia, Michoac\'an, M\'exico;\\       l.loinard@astrosmo.unam.mx}

\author{Claire J.\ Chandler}
\affil{National Radio Astronomy Observatory, P.O.\ Box O, Socorro, NM 87801, USA; cchandle@nrao.edu}

\author{Luis F.\ Rodr\'{\i}guez and Paola D'Alessio}
\affil{Centro de Radiostronom\'{\i}a y Astrof\'{\i}sica, 
       Universidad Nacional Aut\'onoma de M\'exico,\\
       Apartado Postal 72--3 (Xangari), 58089 Morelia, Michoac\'an, M\'exico;\\       l.rodriguez, p.dalessio@astrosmo.unam.mx}

\author{Crystal L.\ Brogan}
\affil{National Radio Astronomy Observatory, 520 Edgemont Road, Charlottesville, VA 22903-2475, USA; cbrogan@nrao.edu}

\author{David J.\ Wilner and  Paul T.P.\ Ho\altaffilmark{1}}
\affil{Harvard-Smithsonian Center for Astrophysics, 60 Garden Street, 
       Cambridge, MA 02138;\\ dwilner, pho@cfa.harvard.edu}

\altaffiltext{1}{Also at: Academia Sinica, Institute of Astronomy and Astrophysics, Taipei 106, Taiwan}

\begin{abstract} 

In this article, we report high-resolution ($\sim$ \msec{0}{1}--\msec{0}{3}),
high-sensitivity ($\sim$ 50--100 $\mu$Jy beam$^{-1}$) Very Large Array
0.7 and 1.3 cm observations of the young stellar system
IRAS~16293--2422 in $\rho-$Ophiuchus. In the 0.7 cm image, component A
to the south-east of the system looks like its usual binary self. In
the new 1.3 cm image, however, component A2 appears to have split into
two sub-components located roughly symmetrically around the original
position of A2. This change of morphology is likely the result of a
recent bipolar ejection, one of the very first such events observed in
a low-mass source.  Also in component A, a marginal detection of 0.7
cm emission associated with the submillimeter component Ab is
reported. If confirmed, this detection would imply that Ab is a
relatively extended dusty structure, where grain coagulation may
already have taken place. With an angular size increasing with frequency, 
and an overall spectra index of 2, the emission from component B to the
north-west of the system is confirmed to be dominated by optically thick 
thermal dust emission associated with a fairly massive, nearly face-on, 
circumstellar disk.  In the central region, however, we find evidence 
for a modest free-free contribution that originates in a structure 
elongated roughly in the east-west direction. We argue that this 
free-free component traces the base of the jet driving the large-scale 
bipolar flow at a position angle of about 110$^\circ$ that has long 
been known to be powered by IRAS~16293--2422.

\end{abstract}


\section{Introduction}

A significant fraction of the stars in the Solar neighborhood are
known to reside and are suspected to have formed in binary or higher
order multiple systems\footnote{The exact fraction may be smaller than
  initially anticipated for very low-mass stars (Lada 2006), but it is
  high for the Solar type objects such as those considered here.}
(Duquennoy \& Mayor 1991). Yet, while the formation of isolated
Solar-type stars is reasonably well understood (Shu, Adams \& Lizano
1987), our comprehension of the formation of multiple stellar systems
remains comparatively limited. Several theoretical mechanisms have
been put forward, but observational constraints on the earliest
evolutionary phases are scarce owing to the difficulty of identifying
and studying the various constituents (disks, jets, etc.) of extremely
young systems (for a recent review on the properties of young
binaries, see Duch\^ene et al.\ 2007). This is because the size scale
of, and separation between, these structures is typically of the order
of a few to a few tens of astronomical units, and so observations with
an angular resolution better than a few tenths of an arcsecond are
needed even in the nearest star-forming regions. Unfortunately, data
with this kind of angular resolution are currently unavailable at the
frequencies (mid-infrared, far-infrared, and submillimeter) best
matched to the energy output of deeply embedded sources. Such
resolutions can only be reached in the optical and near-infrared (but
very enshrouded young stars are not detected at these frequencies),
and at radio wavelengths (0.7 cm $\leq$ $\lambda$ $\leq$ 6 cm) thanks
to the availability of large radio-interferometers.  Most of the
structures found in young stellar systems do emit in the radio domain,
but different mechanisms are at work in different components (thermal
dust emission in accretion disks, thermal bremsstrahlung in jets,
gyrosynchrotron in magnetically active young stars,
etc.). Consequently, a detailed analysis of each radio source found in
a given protostellar system has to be carried out before it can be
identified --or at least associated-- with a specific component.

\section{IRAS~16293--2422}
 
Located in L1689N, a dark cloud in the $\rho-$Ophiuchus star-forming
complex (at $d$ = 120 pc --R.M.\ Torres et al., in prep),
IRAS~16293--2422 is a well-studied very young low-mass protostellar
system (e.g.\ Ceccarelli et al.\ 2000a). It has been suspected to be
multiple since Mundy et al.\ (1986) and Wootten (1989) showed that it
was a double source both at millimeter and centimeter
wavelengths. Soon after, it was also found to power a multi-lobe
outflow system (Mizuno et al.\ 1990), comprising two bipolar and one
monopolar structures. The two bipolar flows are fairly compact, and at
position angles of approximately 60$^\circ$ and 110$^\circ$ (Mizuno et
al.\ 1990, Hirano et al.\ 2001, Castets et al.\ 2001), whereas the
monopolar flow is a parsec-scale, blueshifted feature, located
eastward of IRAS~16293--2422 at a position angle of almost exactly
90$^{\circ}$ (Mizuno et al.\ 1990). The lack of a redshifted
counterpart toward the west is likely related to the location of
IRAS~16293--2422 near the western edge of L1689N. While the two bipolar
flows are now generally accepted to be driven by two distinct
protostellar sources, it is still debated whether the mono-sided
structure is driven by yet another young star in the system, or if it
is an extension of the compact outflow at P.A.\ 110$^\circ$ (Mizuno et
al.\ 1990, Stark et al.\ 2004). In the latter situation, the jet
powering the east-west flows must either have undergone significant
precession, or have been strongly deflected by a condensation of dense
gas along its path.

At the best resolution available at centimeter wavelengths,
IRAS~16293--2422 is resolved into three radio components (Wootten 1989,
Loinard 2002, Chandler et al.\ 2005). Components A1 and A2 --to the
south-east of the system; see Fig.\ 1-- are separated from each other
by about \msec{0}{34}, and from component B, to the north-west, by
about 5$''$. Component B appears to be well resolved in
high-resolution 0.7 cm observations (Rodr\'{\i}guez et al.\ 2005), has a
spectral index of 2--2.5 across the millimeter and centimeter ranges,
and has been interpreted as a nearly face-on, optically thick
accretion disk (Rodr\'{\i}guez et al.\ 2005, Chandler et al.\
2005). It has often been associated with the outflow at P.A.\ $\sim$
110$^\circ$ but significant doubts remain about this issue (Chandler
et al.\ 2005).

Although it has been known for a long time to be at the origin of the
outflow at P.A.\ $\sim$ 60$^\circ$ (Mundy et al.\ 1986, 1992), the
exact nature of the A1/A2 pair remains unclear. Wootten (1989) argued
that A2 is the protostellar source driving the outflow and that A1 is
a shock feature along that flow. This conclusion was based --in part--
on the near-exact alignment with the flow of the segment joining A1 to
A2 at the epoch (1986-1987) of the observations published by Wootten
(1989). In that scheme, A1 could either be a knot of ionized gas
ejected by A2 some time in the past, or the result of the impact of
the jet powered by A2 onto dense circumstellar material. In the former
case, one would expect A1 to move steadily away from A2 along the
direction of the flow, whereas in the latter, no significant motions
are expected between A1 and A2. Multi-epoch radio observations with
sufficient angular resolution (better than about \msec{0}{2}) to
resolve A1 and A2, however, revealed neither of these behaviors
(Loinard 2002, Chandler et al.\ 2005). Instead, they showed that the
separation between A1 and A2 has remained constant at about
\msec{0}{34} in the last 15 years, whereas the position angle between
them has changed from less than 50$^\circ$ in the late 1980s to about
85$^\circ$ in 2003 (Loinard 2002, Chandler et al.\ 2005). This
definitely rules out the possibility that A1 is an ejecta from A2,
because --as mentioned earlier-- one would then have expected constant
position angle and increasing separation rather than the opposite. The
possibility that A1 is the result of the interaction between a jet
powered by A2 and circumstellar material could still be retained if
that jet precessed strongly between the late 1980s and 2003, and were
now oriented almost exactly east-west (at P.A.\ $\sim$ 85$^{\circ}$).
Recent observations, however, have shown that the 0.7 cm radio emission
from A2 (that partly traces free-free radiation from the inner part of
the jet) is elongated at P.A.\ $\sim$ 60$^\circ$, very similar to the
orientation of the large-scale molecular flow driven from within
component A. Thus, it appears that A2 {\em does} drive the large-scale
flow at P.A.\ $\sim$ 60$^\circ$, and that A1 is {\em not} a shock
feature along that flow.

The simplest interpretation of the relative motion between A1 and A2
(constant separation and linearly increasing position angle) is that
they are two protostellar sources in Keplerian orbit (Loinard
2002). In this hypothesis, the total mass of the system can be
estimated from Kepler's third law. Assuming a face-on circular orbit,
the mass of the system implied by the motions is 2.7(d/120 pc)$^3$
\Msun\ (where $d$ is the distance --Chandler et al.\ 2005). For an
elliptical orbit, a somewhat smaller value would be permitted, but the
total mass has to be larger than 1.4 (d/120 pc)$^3$ \Msun\ for the
system to remain bound. Moreover, this minimum value is only attained
for an extremely eccentric orbit observed exactly at periastron. For
less special circumstances, a mass larger than 2 to 3(d/120 pc)$^3$
\Msun\ is required. Finally, an analysis of the absolute proper
motions (Loinard 2002, Chandler et al.\ 2005) shows that, in this
orbital motion scenario, the center of mass would have to be very
close to A2, so that A2 would have to be significantly more massive
than A1. The A1/A2 pair would then be a binary system made of a
relatively low-mass protostar ($M$ $\lesssim$ 0.5 \Msun) in orbit
around a somewhat more massive one ($M$ $\gtrsim$ 2 \Msun). That
binary system would, in turn, be orbited by the protostar associated
with component B, and IRAS~16293--2422 would be a hierarchical triple
system. Note that if A1 and A2 indeed form a binary system, then the
position angle between them should keep increasing monotonically with
time. Furthermore, if observations with a quality similar to those
used by Loinard (2002) and Chandler et al.\ (2005) are obtained
regularly in the coming decade, it will become possible to fit a
Keplerian orbit to the data, and determine accurately the mass of
both components.

An alternative to this orbital motion scenario was recently formulated
by Chandler et al.\ (2005). In that proposal, A1 is interpreted as a
shock resulting from the impact of a strongly precessing (or wobbling)
jet driven by a third --as-yet undetected-- protostar in the
system. Of course, in this alternative scheme, the change of position
angle with time should decelerate, and eventually reverse its course
because the jet must oscillate around an equilibrium value. Note that
the required fast rate of precession or wobbling (about 40$^\circ$ in
less than twenty years) has never been observed before, and would
require the new protostar to be a member of a very tight binary. Given
the overall morphology of the system, the most likely situation would
be for this new member to be a close companion of A2 --with a
separation smaller than 22 AU (\msec{0}{18} if d = 120 pc --Chandler
et al.\ 2005). It is interesting to point out that, in that
alternative scheme, there would also be two protostellar sources in
the A component and, including component B, IRAS~16293--2422 would
again be a hierarchical triple system. Thus, the two scenarios put
forth so far call for the existence of three protostellar sources in
IRAS~16293--2422. Given that three outflow systems are known to be
powered from within IRAS~16293--2422 (see above), it is, of course,
tempting to associate each one of the sources with one of the flow
features. So far, however, the only secure association is that between
A2 and the flow at P.A.\ $\sim$ 60$^\circ$.

In recent high-resolution, submillimeter observations (Chandler et
al.\ 2005), IRAS~16293--2422 was also found to be a triple source (Aa,
Ab, and B --see Fig.\ 1), but the exact correspondence between the
radio and submillimeter sources in the A component is not immediately
obvious\footnote{The situation is clear for component B, where the
position of the submillimeter and radio sources are almost identical,
and the submillimeter flux is in excellent agreement with that
expected from the extrapolation of the centimeter data.}. Component Aa
is located just about halfway between A1 and A2 (Fig.\ 1) and it is
unclear if it is another protostar in the system, or a blend of
roughly equal amounts of emission associated with A1 and A2.  If it
was a new source, its position roughly \msec{0}{2} from A2 would, of
course, make it a potential candidate for the companion of A2 driving
the precessing jet that impacts circumstellar material at
A1. Component Ab, on the other hand, is located about \msec{0}{6} to
the north-east of component A, and is not detected at any other
wavelengths. Since the mechanism at work is almost certainly dust
thermal emission, Ab must be a dense dusty structure, and Chandler et
al.\ (2005) argued that it may be associated with a fourth protostar
in the system. The absence of centimeter emission, however, shows that
it is not associated with a strong ionized wind. Moreover, for thermal
dust emission, the spectral energy distribution is expected to be a
power law with an index $\alpha$ between 2 and 4. From the 0.5 Jy flux
detected at 305 GHz, one would expect a 0.7 cm flux of about 10 mJy if
the spectral index is 2, and 0.2 mJy if $\alpha$ is 4. A 10 mJy source
would have been easily detected in the 0.7 cm observations reported by
Rodr\'{\i}guez et al.\ (2005) unless the emission were fairly extended
($\gtrsim$ 0.5$''$) and significantly filtered out.  Only a very
marginal detection (2$\sigma$) would have been obtained in the second
case ($\alpha$ = 4), but the source would have been detected above
5$\sigma$ for any spectral index smaller than 3.5 (again, unless the
emission were extended). Thus, the non-detection of Ab in the 0.7 cm
data reported by Rodr\'{\i}guez et al.\ (2005) implies either that the
emission is fairly extended, or that the spectral index is larger than
3.5, and the dust consequently largely unprocessed (or a combination
of both). Together with the lack of free-free emission, these
characteristics argue in favor of either an extremely young
protostar, or even a starless condensation.

IRAS~16293--2422 has long been known to exhibit a very active
chemistry (e.g.\ van Dishoeck et al.\ 1995, Cazaux et al.\ 2003), 
and is one of the sources where the hot corinos (Ceccarelli et al.\ 
2007) that have been argued to trace the earliest stages of low-mass 
star-formation have first been identified. The earliest observations 
(e.g.\ Ceccarelli et al.\ 2000b and references therein) did not 
resolve the various components of IRAS~16293--2422, but several recent 
high-resolution interferometric millimeter observations (Kuan et
al.\ 2004, Bottinelli et al.\ 2004, Chandler et al.\ 2005) have
demonstrated that the organic and highly deuterated molecules
considered signposts of hot corinos exist in both component A and
component B. It appears, therefore, likely that each component harbors
at least one very young protostar.

\bigskip

In summary, although it has been extensively studied and several
characteristics of IRAS~16293--2422 are now well-established, many
other questions remain open about the very nature of several of the
sources in the system, and their exact relation to the large-scale
outflows in the region. It is important to address these issues,
because IRAS~16293--2422 is a rare example of a very young multiple
system, that offers a unique opportunity to study the earliest stages
of the formation of multiple systems.

\section{Observations} 

In this article, we will present and analyze
two new, high-resolution, high-sensitivity, radio continuum
observations of IRAS~16293--2422 obtained at 0.7 and 1.3 cm (Tab.\
1). The data were collected with the {Very Large Array} (VLA) of the
{National Radio Astronomy Observatory} (NRAO\footnote{NRAO is a
facility of the National Science Foundation operated under cooperative
agreement by Associated Universities, Inc.}) in the most extended (A)
and second most extended (B) configurations of the array at 1.3 cm and
0.7 cm, respectively. The standard VLA continuum frequency setups were
used: two frequencies (22.4851 and 22.4351 GHz at 1.3 cm, and 43.3149
and 43.3649 GHz at 0.7 cm) were observed simultaneously, and in both
circular polarizations, with 50 MHz of bandwidth each. The absolute
flux density was set using observations of 3C 286. For improved flux
accuracy, we did not assume 3C 286 to be a point source, but instead
used a model image provided by NRAO. The phase calibrator was PKS
J1625--2527 whose absolute position is expected to be accurate to
about 2 milli-arcseconds.  The fast-switching technique --that
consists of rapidly alternating observations of the source and the
phase calibrator with a cycle time of 2 minutes-- was used for both
observations; this ensured optimal imaging fidelity. The 0.7 cm data
were restored with natural weighting of the visibilities, whereas the
1.3 cm data were imaged with a robust weighting scheme intermediate
between uniform and natural (see Tab.\ 1 for the corresponding 
synthesized beams).

To complement these new data, we will also make use of observations
previously published in the literature. In particular, we will use the
recent 0.7 cm observations obtained in 2003.47 and reported by
Rodr\'{\i}guez et al.\ (2005), and the 3.6 cm high-resolution
observations obtained in 2003.65, and reported by Chandler et al.\
(2005). For these two datasets, new images were made from the
calibrated visibilities. The 0.7 cm data from 2003.47 were 
initially imaged with natural weighting of the visibilities. 
For the study of component A (which is faint at 0.7 cm) as well
as for accurate comparison with the new 1.3 cm data, this image 
was also smoothed to the resolution of the 1.3 cm image (Tab.\ 1)
The 3.6 cm data were imaged with a robust weighting scheme 
intermediate between natural and uniform. 

Our new B-array 0.7 cm observations were found to produce fluxes
systematically 30\% lower than the older A-array observations. 
This was originally identified by comparing the total flux of 
component B (which has never been found to be variable in any 
previous observations) in the two datasets. A similar difference 
was found when comparing the total flux corresponding to similar 
baselines and hour angles in our A- and B-configuration observations. 
This effect clearly cannot be due to missing flux related extended
emission, since this would have produced more flux in the 
B-configuration observations than in the A-array data, rather
than the observed opposite. Instead, the flux deficit is likely 
related to the relatively poorer weather conditions during our B-array 
observations. To account for this effect, the new B-configuration
0.7 cm data were multiplied by 1.3 before imaging. The only instance 
when this will be important will be in Sect.\ 4.3, when we estimate 
the flux of the possible 0.7 cm counterpart of Ab.

\section{Results and discussion}

\subsection{Component A}

In our new 0.7 cm observation (Fig.\ 2b), component A looks like its
usual binary self, with sub-components A1 and A2 clearly resolved. In
the new 1.3 cm image, however, it has clearly become triple (Fig.\ 2a,
particularly the inset). This is the first time that component A
appears triple rather than double in a radio image. While the
easternmost source is clearly A1, the relation between the two western
sources and A2 is {\it a priori} quite unclear. Since they are
identified here for the first time, we have very little information on
the spectral energy distributions of the two western sources. The only
statement that can safely be made on their spectral properties is that
they are both continuum (rather than spectral) features because they
are detected with identical fluxes at both observed frequencies
(22.4851 and 22.4351 GHz --see Sect.\ 2). It is also noteworthy that
the total flux of component A in our new 1.3 cm image (5.2 $\pm$ 0.3
mJy) is comparable to the 1.3 cm flux of component A in all previous
observations (see Fig.\ 7, left panel in Chandler et al.\ 2005). Thus,
while the {\em morphology} of A2 has undergone significant changes,
its {\em flux} has not. This suggests, in particular, that the current
overall spectral index of component A is similar to that found in
previous observations. In the centimeter regime, that spectral index
is about 0.53 (see Fig.\ 2, left panel in Chandler et al.\ 2005)
suggesting that free-free is the dominant emission process.

To examine further the relation between the three sources associated
with component A in our new 1.3 cm image and the usual sub-components
A1 and A2, it is useful to superimpose our new 1.3 cm map onto one of
the previous high resolution radio images of component A. For this
purpose, we will use the smoothed version of the A-array 0.7 cm image
(Sect.\ 3), because it has been obtained relatively shortly before the 
1.3 cm data presented here, and because it provides a good compromise 
between angular resolution and sensitivity. Since the absolute positions
of the various components of IRAS~16293--2422 change with time 
(Loinard 2002, Chandler et al. 2005), one has to properly register the 
images before the superposition. The absolute proper motion of A2 is
known to be very similar to that of component B (Chandler et al.\ 2005).
From the four recent observations considered in this paper, the proper
motion of component B can be calculated to be 

\[ \mu_\alpha \cos \delta = -7.6 \pm 1.2 \mbox{~mas yr$^{-1}$~~~~;~~~~~} \mu_\delta = -25.8 \pm 1.1 \mbox{~mas yr$^{-1}$} \]

\noindent The latter value is somewhat larger than those quoted by
Loinard et al.\ (2002) and Chandler et al.\ (2005). The reasons for
this discrepancy are not entirely clear, but may be related to the
fact that the older data used by Loinard et al.\ (2002) and Chandler 
et al.\ (2005) are of lower quality than the more recent observations
considered here. In any case, we will use the values quoted above, 
since they are more appropriate for the comparisons that will
be made in this paper between recent observations. Given that the 0.7 
cm image was obtained 2.64 years before the present 1.3 cm data, it was 
shifted by $-$20.1 mas and $-$68.0 mas in right ascension and declination, 
respectively, before the superposition.  The result of the comparison 
(Fig.\ 3) confirms that the easternmost 1.3 cm component is nearly exactly 
coincident with the 0.7 cm source A1, and can indeed be identified with 
it. Neither of the two western 1.3 cm sources, however, appears to coincide 
with source A2. Instead, they are located roughly symmetrically on each 
side of A2, and we shall refer to them as A2$\alpha$ and A2$\beta$ in the 
rest of the paper (Fig.\ 2). The line joining A2$\alpha$, A2, and A2$\beta$
(thick dotted line on Fig.\ 3) is at a position angle of about
62$^\circ$, remarkably similar to the direction of the large-scale
flow known to be driven by A2. This strongly suggests that A2$\alpha$
and A2$\beta$ are the result of a recent bipolar ejection by the
protostar within A2, the position of which is likely near the cross in
Fig.\ 3. While bipolar ejections are not uncommon in young stellar
objects (e.g. Mart\'{\i} et al.\ 1995, Curiel et al.\ 2006), they
rarely lead to such dramatic morphological changes.  In particular, in
our case, it seems that very little emission is left at the very
position of the protostar, the emission concentrating instead in the
two bipolar lobes. This usually does not happen in bipolar ejections
by protostars. Also, while bipolar ejections have been observed from
relatively massive young stellar system, this is --to our knowledge--
the first time that it is seen in such a low-mass source.  It is
interesting to note that a situation almost exactly opposite of that
reported here was recently found in a radio source in Orion. Source
``n'', which had been bipolar in all previous VLA observations, was
indeed found to be single in the 2007 VLA image to be published by L.\
G\'omez et al.\ (in prep).

A2$\alpha$ and A2$\beta$ were clearly not present in previous radio
images so if they are indeed ejecta, they must be very recent
ones. Water masers associated with component A are found to expand at
a velocity projected on the plane of the sky of 40--65 km $^{-1}$
(Wootten et al.\ 1999). If this value can be taken as representative
of the projected velocity of the jet powered by A2, then it would have
taken just about a year for A2$\alpha$ and A2$\beta$ to reach the
present position. This is consistent with their absence from any
previous observation. Note, also, that if A2$\alpha$ and A2$\beta$ are
ejecta moving at about 50 km s $^{-1}$ away from the protostar
associated with A2, then their proper motions should be easily
detectable. An observation of IRAS~16293--2422 in the upcoming A
configuration of the VLA (around 2007.8), should show A2$\alpha$
and A2$\beta$ displaced from their current position by about
\msec{0}{1}.

\subsection{Relative motion between A1 and A2}

The relative motion between A1 and A2 in the last 15 years or so has
been investigated in detail by Loinard (2002) and Chandler et al.\
(2005). As mentioned earlier, these studies have led to the conclusion
that the separation between A1 and A2 has remained constant at about
\msec{0}{34}, whereas their relative position angle has changed by
about 40$^\circ$ between the late 1980s and 2003. The new 0.7 cm
observation reported here can be used to further monitor the evolution
of the relative position of A1 and A2. The value of the position angle
found in that observation (82 $\pm$ 3$^\circ$ --Fig.\ 4) is similar to
those reported for the 2003 observations by Chandler et al.\
(2005). Obtaining information from the new 1.3 cm observation is
somewhat more difficult because of the change of structure undergone
by component A2. In Fig.\ 4, we report the separation and position
angle between A1 and both A2$\alpha$ and A2$\beta$. As discussed
earlier, however, the protostar in A2 is unlikely to be coincident
with either of these two sources. Instead, it must be near the cross
in Fig.\ 3, a location roughly equivalent to the mean of the positions
of A2$\alpha$ and A2$\beta$. The red circles in Fig.\ 4 indicate the
separation and position angle between A1 and the cross in Fig.\
3. Taken together, these new observations suggest that the separation
between A1 and the protostar associated with A2 has remained constant
around \msec{0}{34} in the last few years, whereas their relative
position angle is between 80 and 90$^\circ$. Unfortunately, this does
not shed much new light on the relative motion between A1 and A2. In
particular it is insufficient to discriminate between the orbital path
scenario (where one would expect the position angle to keep increasing
linearly) and the precessing/wobbling scheme (where the change of
position angle should decelerate and eventually reverse its course).
More observations in the coming decade will be needed to settle this
issue.

\subsection{0.7 cm emission from component Ab?}

We mentioned in Sect.\ 1 that the submillimeter source Ab, located
\msec{0}{6} to the north-east of Component A has, to date, never been
detected at other wavelengths. To further examine this issue, we
searched both of our new images for a counterpart of Aa. While there
is clearly no detection in the present 1.3 cm data, a possible source
is detected at the 4$\sigma$ level in our new B-configuration 0.7 cm
dataset (Fig.\ 5). The best 2-dimensional Gaussian fit to this
structure yields a peak flux density of about 0.76 mJy beam$^{-1}$, and
an integrated flux of 1.8 $\pm$ 0.6 mJy. Note that the noise level in 
the 0.7 cm A array observations published by Rodr\'{\i}guez et al.\ 
(2005) was 0.1 mJy beam$^{-1}$, so this source should have been detected 
in that dataset also if it had been present. The fact that it was not, 
implies either that the present marginal detection is not real, or that 
the 0.7 cm emission associated with Ab is extended, and was more heavily 
filtered out in the A array observations published by Rodr\'{\i}guez et 
al.\ (2005) than in the present B array data. Deep, C-configuration VLA
observations at 0.7 cm ought to settle this issue. If the marginal
detection reported here is real, then the spectral index of the
emission associated with component Ab must be smaller than about 3
independently of the extent of the emission.  Thus, component Ab would
be a relatively extended structure containing somewhat processed
dust. Combined with the lack of centimeter emission tracing winds,
this would favor a scenario where Ab is a starless condensation rather
than a protostar.


\subsection{Component B: disk and jet combined}

Component B is clearly resolved in our new 1.3 cm observations (Fig.\
6.c), and its morphology is very similar to that in the A-configuration
0.7 cm image smoothed to a similar resolution (Fig.\ 6.a). It is slightly 
more compact, however, at 1.3 cm (\msec{0}{19} $\times$ \msec{0}{15}) than
at 0.7 cm (\msec{0}{21} $\times$ \msec{0}{20}). This is consistent
with the trend noticed by Chandler et al.\ (2005) between archival 1.3
cm and 3.6 cm data --in the 3.6 cm image shown in Fig.\ 1, the source 
is found to have a deconvolved size of \msec{0}{16} $\times$ 
\msec{0}{10}-- and suggests that the (linear) angular size of the 
emission increases roughly as $\nu^{+0.3}$. This increase of the angular 
size with frequency is opposite to what is expected from optically thick 
free-free emission. Combined with the measured spectral index in the 
centimeter regime of about 2 (Chandler et al.\ 2005), this strongly 
suggests that the dominant radiation mechanism is optically thick thermal 
dust emission as proposed before (Rodr\'{\i}guez et al.\ 2005, Chandler 
et al.\ 2005). Clearly, to make the emission optically thick, the 
circumstellar disk associated with component B must be quite dense and 
massive (Chandler et al.\ 2005). Since the emission is well-resolved in 
our data, we can investigate if and how the spectral index depends on 
radius. We find that the spectral index in the very inner region (at R $<$
\msec{0}{1}) is 1.4 $\pm$ 0.2, whereas it is 2 $\pm$ 0.2 in the region
\msec{0}{1} $<$ R $<$ \msec{0}{2} (Fig.\ 6.d). The spectral index outside 
of R = \msec{0}{2} becomes difficult to calculate reliably because the 
emission fades quickly, but it is clearly larger than 2. The lower value 
of the spectral index near the center of component B suggests that the 
optically thick thermal dust emission in these regions coexists with a 
lower spectral index component. The most likely candidate for this 
additional component would be free-free radiation from the base of an 
ionized wind. Such a wind would indeed be expected to exist in an 
actively-accreting protostar like that believed to reside at the center 
of component B.

It is interesting in this context, to compare the structure of the
inner and outer regions of component B. The image most suitable for
that purpose is the full-resolution 0.7 cm image obtained in the A 
configuration of the VLA (Fig.\ 6.b), because it provides the
best compromise between sensitivity and angular resolution. It is
noteworthy in that image that the outer isophotes are elongated
roughly in the north-south direction (at a position angle marginally
positive), whereas the inner isophotes are elongated roughly in the
east-west direction (at P.A.\ $\sim$ 110$^\circ$) very similar to that
of one of the large-scale outflows known to originate from within
IRAS~16293--2422. This would be consistent with the idea that
component B is at the origin of that flow; the radio source associated
with it being the superposition of a compact thermal jet at P.A.\
$\sim$ 110$^\circ$, and a somewhat extended disk roughly perpendicular
to the jet. It is puzzling in that scenario, however, that no high-velocity
emission is detected towards component B in any molecular transition
(Chandler et al.\ 2005). If a thermal jet is associated with the
protostar near the center of component B, one would indeed expect to
find entrained high velocity gas associated with it. Future,
high-resolution, millimeter and submillimeter spectral observations
might help resolve this puzzle.

\section{Conclusions and perspectives}

The new observations presented in this paper provide a number of
interesting clues on the nature and properties of the various radio
sources in IRAS~16293--2422.

\begin{enumerate}

\item Component A2 is definitely confirmed to be at the origin of the
compact bipolar flow at P.A.\ $\sim$ 60$^\circ$. Indeed, in the new
1.3 cm image, A2 has undergone a dramatic morphological change, being
now composed of two emission peaks distributed roughly symmetrically
around A2 at a position angle of almost exactly 60$^\circ$. We argue
that this change reflects a recent bipolar ejection, an interpretation 
that will be confirmed if forthcoming observations reveal the
appropriate proper motions.

\item Unfortunately, the new 0.7 and 1.3 cm data do not allow us to
characterize much better the relative motion between A1 and A2.  It
remains unclear, therefore, whether the changes in relative position
observed in the last two decades reflect an orbital motion, or the
precession or wobbling of a jet. It will be important to settle that
issue to decide where the third star in the system is, and if that
star might be at the origin of the third outflow structure known to
exist in IRAS~16293--2422. Further observations in the next ten years
or so ought to provide a definite answer.

\item A marginal detection of 0.7 cm emission near the position of the
submillimeter source Ab is reported. It would be very important to
further characterize the very nature of this source, and its relation
to the other members of the system. If the present marginal detection
is real, it should be easily confirmed by future deep 0.7 cm
observations in the C configuration of the VLA.

\item The radio emission from component B at the northwest of the
system is confirmed to be dominated by optically thick thermal dust
emission, suggesting that the radio source in that component traces an
optically thick accretion disk. At the very center of that structure,
however, the lower observed value of the spectral index suggests the
existence of a modest contribution from a thermal jet.  The
orientation of the inner 0.7 cm isophotes further suggests that this
free-free component traces the base of the compact bipolar flow at
P.A.\ $\sim$ 110$^\circ$. This association would appear reasonable
since the flow at P.A.\ $\sim$ 110$^\circ$ has a dynamical age of only
a few thousand years (Mizuno et al.\ 1990), and component B is known
to harbor organic and deuterated molecular species expected to be
present only in the youngest protostellar sources (Cazaux et al.\
2003, Kuan et al.\ 2004, Bottinelli et al.\ 2004, Chandler et al.\
2005).

\end{enumerate}

\acknowledgements
L.L., L.F.R., and P.A.\ acknowledge the financial support of DGAPA,
UNAM and CONACyT, M\'exico.  D.J.W. acknowledges partial support from 
NASA Origins of Solar Systems Program Grant NAG5-11777.

\clearpage

\begin{table}[!t] 
\caption{Observing logs} 
\vspace{0.4cm} 
\centering
\begin{tabular}{lcrclcll} 
\hline \hline
Date & Project & \multicolumn{1}{c}{$\nu$} & Conf. & Weighting/ & 
Synthesized beam & Figs & Refs\tablenotemark{a}\\%
     &         & (GHz)     & & Smoothing\\%
\hline
2003.47 & AL592 & 43.2299 & A & Nat. & \sbeamp{0}{09}{0}{05}{1.8} & 6b & 1,2 \\%
2003.47 & AL592 & 43.2299 & A & Nat./Smo. & \sbeamm{0}{13}{0}{06}{0.5} & 3, 6a & 3 \\%
2003.65 & AL589 & 8.4601  & A & Rob. & \sbeamp{0}{39}{0}{19}{6.8} & 1 & 2 \\%
2005.20 & AC778 & 43.2299 & B & Nat. & \sbeamm{0}{30}{0}{17}{1.9} & 2b, 5 & 3\\%
2006.11 & AL672 & 22.4601 & A & Rob. & \sbeamm{0}{13}{0}{06}{0.5} & 2a, 3, 6c & 3\\%
\hline \hline \\
\end{tabular}
\tablenotetext{a}{1=Rodr\'{\i}guez et al.\ (2005); 2=Chandler et al.\ (2005); 3=This work}
\end{table}

\clearpage
\begin{figure}[!t]
\centerline{\includegraphics[height=0.6\textwidth,angle=-90]{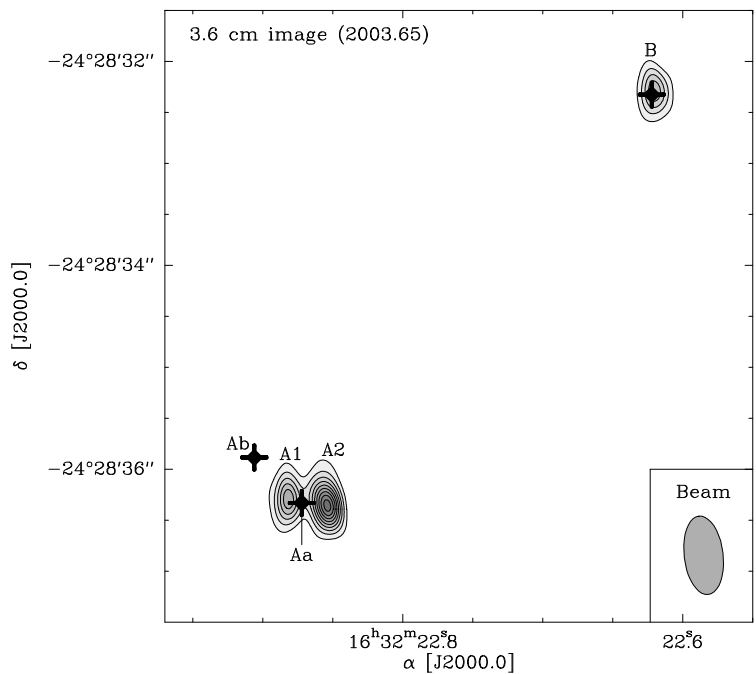}}
\caption{3.6 cm VLA A-array image of IRAS~16293--2422 obtained in
2003.65 (Chandler et al.\ 2005) and restored with a weighting scheme
intermediate between natural and uniform. The first contour and the
contour interval are 0.1 mJy beam$^{-1}$, while the noise level is
0.02 mJy beam$^{-1}$. The synthesized beam, shown at the bottom-right
corner, is \sbeamp{0}{39}{0}{19}{6.8}. The radio components A1, A2, 
and B are labeled, and the position of the submillimeter sources Aa, 
Ab, and B are indicated. The errors on the submillimeter source positions 
are significantly smaller than the crosses indicating their locations.} 
\end{figure}

\clearpage
\begin{figure*}[!t]
\centerline{\includegraphics[width=0.85\textwidth,angle=-90]{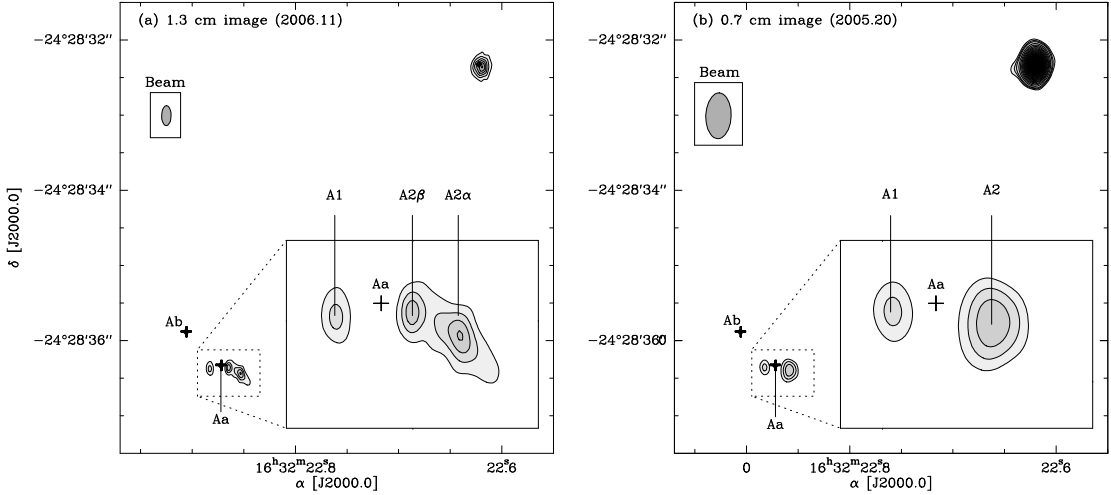}}
\caption{(a) 1.3 cm continuum observation of IRAS~16293--2422 obtained
in 2006.11 restored with a weighting scheme of the visibilities
intermediate between natural and uniform. The first contour is at 0.2
mJy beam$^{-1}$ and the contour step is 0.3 mJy beam$^{-1}$, while the
noise is 0.04 mJy beam$^{-1}$. The synthesized beam, shown near the
top-left corner, is \sbeamm{0}{13}{0}{06}{0.5}. (b) 0.7 cm image of 
IRAS~16293--2422 obtained in 2005.20 in the B configuration of the
VLA when restored with natural weighting. The first contour is at 1.5 
mJy beam$^{-1}$ and the contour step is 0.3 mJy beam$^{-1}$, while the 
noise is 0.18 mJy beam$^{-1}$. The synthesized beam, shown near the
top-left corner, is \sbeamm{0}{30}{0}{17}{1.9}. In both panels, the 
positions of the submillimeter sources Aa and Ab are indicated,
and insets provide zooms on component A. Note that component A is
clearly triple in the 1.3 cm image, while it remained double in the
0.7 cm map.} \end{figure*}

\clearpage
\begin{figure}[!t]
\centerline{\includegraphics[width=0.35\textwidth,angle=270]{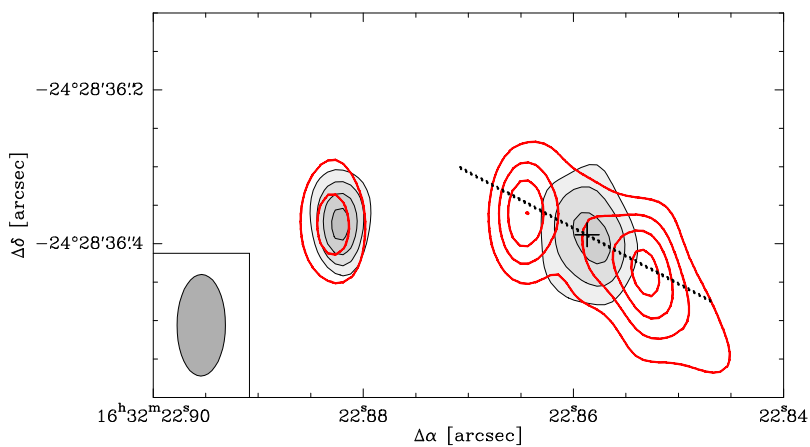}}
\caption{Comparison between the smoothed 0.7 cm image of component A 
obtained in 2003 (black contours and grey scale) and our new 1.3 cm image 
(red contours). The angular resolution of the two images is identical
(\sbeamm{0}{13}{0}{06}{0.5}) and is shown at the bottom-left. Translational 
shifts have been applied to account for the overall proper motion of the 
region. The first contour and the contour interval for the 1.3 cm image 
are the same as in Fig.\ 2; they are at 0.6 and 0.15 mJy beam$^{-1}$, 
respectively, for the 0.7 cm image. The dotted line is at a position 
angle of 62$^\circ$}
\end{figure}

\clearpage
\begin{figure*}[!t]
\centerline{\includegraphics[width=0.4\textwidth,angle=270]{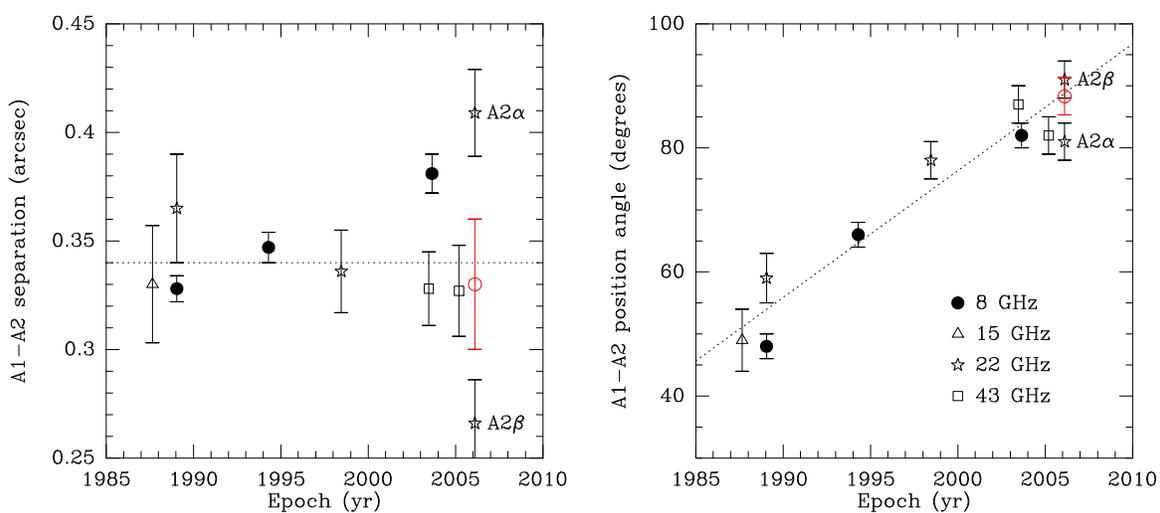}}
\caption{Evolution of the separation (left) and relative position
angle (right) of the A1/A2 pair. For the 1.3 cm data obtained in
2006.11, three points are shown: in black the parameters for
A2$\alpha$ and A2$\beta$, and in red those corresponding to the
position were the protostar is inferred to be. The dotted line in the
left panel shows a constant separation of \msec{0}{34}, whereas the
dotted line in the right panel shows the best linear fit to all the
data points but those corresponding to the 2006.11 image.}
\end{figure*}

\clearpage
\begin{figure}[!t]
\centerline{\includegraphics[width=0.35\textwidth,angle=270]{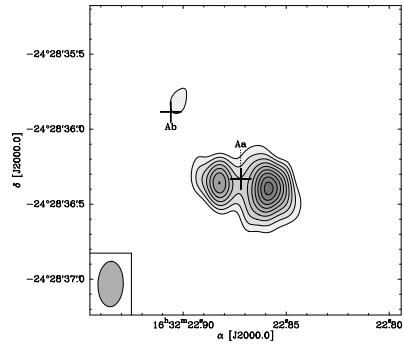}}
\caption{0.7 cm image of component A. The first contour is at 0.7 mJy
beam$^{-1}$, and the contour interval is 0.2 mJy beam$^{-1}$, while
the noise is 0.18 mJy beam$^{-1}$. The synthesized beam, shown at the
bottom-left corner is \sbeamm{0}{30}{0}{17}{1.9}. Note the existence 
of a positive signal near the position of Ab.} 
\end{figure}

\clearpage
\begin{figure*}[!t]
\centerline{\includegraphics[width=0.8\textwidth,angle=0]{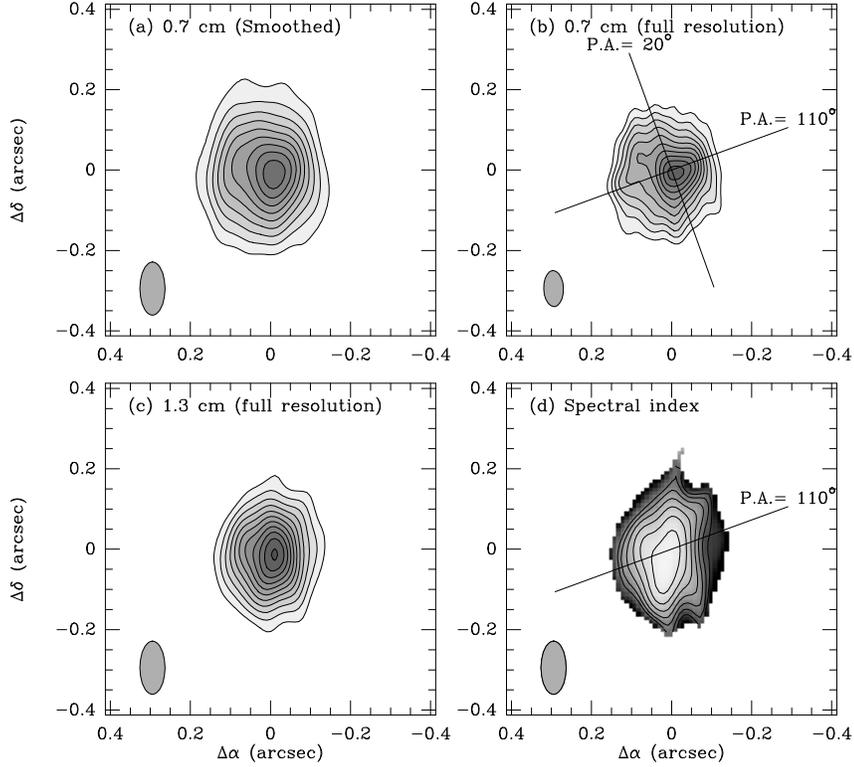}}
\caption{Comparison between high angular resolution images of
component B in IRAS~16293--2422 obtained at 0.7 cm (panels a--b) and
1.3 cm (panel c). In panel (b), we show the full-resolution 0.7 cm
combined A+B configuration image. The first contour is at 0.6 mJy
beam$^{-1}$ and the contour interval is 0.2 mJy beam$^{-1}$, while the
noise level is 0.12 mJy beam$^{-1}$. The synthesized beam, shown at
the bottom-right corner is \sbeamp{0}{09}{0}{05}{1.8}. In panel
(c) we show the full resolution 1.3 cm A configuration image. The
first contour is at 0.2 mJy beam$^{-1}$ and the contour interval is
0.15 mJy beam$^{-1}$. In panel (a) we show the 0.7 cm image smoothed
to the resolution of the 1.3 cm image. The first contour is at 0.8 mJy
beam$^{-1}$ and the contour interval is 0.4 mJy beam$^{-1}$. Finally,
in panel (d), we show the spectral index between 0.7 and 1.3 cm. The
grey scale goes from 1 (white) to 3 (black). The coutours are at 1.4,
1.6, 1.8, 2, 2.2, 2.4, and 2.6. The resolution in panels (a), (c) and 
(d), shown at the bottom-left of each panel is \sbeamm{0}{13}{0}{06}{0.5}}.
\end{figure*}
\clearpage
\end{document}